\documentclass[twocolumn,aps,superscriptaddress,showpacs,nofootinbib,floatfix]{revtex4}

\usepackage{epsfig,bm,feynmf}

\usepackage{graphics}

\usepackage[normalem]{ulem}  
\usepackage[dvips]{color} 

\renewcommand\sout{\bgroup \color{red} \ULdepth=-.5ex \ULset}

\begin{document}



\title{Charmonium production from nonequilibrium charm and anticharm quarks
in quark-gluon plasma}


\author{Taesoo Song}\email{songtsoo@yonsei.ac.kr}
\affiliation{Cyclotron Institute, Texas A$\&$M University, College Station, Texas 77843-3366, USA}
\author{Kyong Chol Han}\email{khan@comp.tamu.edu}
\affiliation{Cyclotron Institute and Department of Physics and Astronomy, Texas A$\&$M University, College Station, Texas 77843-3366, USA}
\author{Che Ming Ko}\email{ko@comp.tamu.edu}
\affiliation{Cyclotron Institute and Department of Physics and Astronomy, Texas A$\&$M University, College Station, Texas 77843-3366, USA}


\begin{abstract}
Parameterizing the charm and anticharm quark momentum distributions by the Tsallis distribution, we study the nonequilibrium effect on the charmonium production rate in a quark-gluon plasma up to the next-to-leading order in perturbative QCD. We find that nonequilibrium charm and anticharm quarks suppress the charmonium production rate compared to that from equilibrated ones. We further show that the suppression factor calculated with the charm quark relaxation time, which has been frequently used in the literature, is close to our results.
\end{abstract}

\pacs{25:75.Cj} \keywords{}

\maketitle


\section{introduction}

Since the suggestion of using charmonium suppression as a possible signature for the formation of a quark-gluon plasma (QGP) in relativistic heavy-ion collisions~\cite{Matsui:1986dk}, there have been many theoretical~\cite{Vogt:1999cu,Zhang:2000nc,Zhang:2002ug,Yan:2006ve,Zhao:2007hh,Song:2010ix,Song:2010er,Zhao:2011cv,Song:2011xi}  and experimental~\cite{Alessandro:2004ap,Adare:2006ns,:2010px,Silvestre:2011ei,MartinezGarcia:2011nf} studies on charmonium production in heavy ion collisions at energies available from the Super Proton Synchrotron (SPS), the Relativistic Heavy Ion Collider (RHIC), and the Large Hadron Collider (LHC). Although the collision energy spans two orders of magnitude from SPS to LHC, the observed suppression of charmonium production in these collisions was found to be similar. A possible explanation for this surprising result is that the expected larger suppression of charmonium production with increasing collision energy is compensated by the increasing number of regenerated charmonia from initially produced charm and anticharm quarks in QGP. This is because the number of regenerated charmonia is quadratically proportional to \cite{Andronic:2006ky} while that of directly produced charmonia depends linearly on the number of charm quarks~\cite{Andronic:2006ky,Adare:2006kf,Adare:2010de}. Studies based on phenomenological models indeed show that although the fraction of regenerated charmonium is compatible to that of directly produced one in central Au+Au collisions at RHIC, it becomes dominant in central Pb+Pb collisions at LHC~\cite{Yan:2006ve,Zhao:2007hh,Song:2010ix,Zhao:2011cv,Song:2011xi} as more charm and anticharm quarks are produced.

The charmonia production rate in QGP depends on the charm and anticharm quark distributions in both momentum and coordinate spaces~\cite{Greco:2003vf}. Presently it is not clear to what degree charm and anticharm quarks, which are produced from initial hard collisions with a power law distribution, are thermalized in relativistic heavy-ion collisions. It seems that charm and anticharm quarks of low transverse momentum are close to while those of intermediate and high transverse momenta are far from thermal equilibrium~\cite{Song:2011kw}. In Refs.~\cite{Zhao:2007hh,Song:2010ix,Song:2010er,Zhao:2011cv,Song:2011xi}, the charm quark nonequilibrium effect on charmonia regeneration in QGP is included by multiplying the expected equilibrium charmonium number by a relaxation factor $R(\tau)=1-\exp[-(\tau-\tau_0)/\tau_{\rm rel}]$ with $\tau_{\rm rel}$ being the relaxation time of charm quarks in QGP. Since the latter has a value $\tau_{\rm rel}\sim 4~{\rm fm}/c$, the regeneration contribution to charmonium production is suppressed by a factor of 0.53, 0.68 and 0.8 in heavy ion collisions at SPS, RHIC, and LHC, respectively, when the corresponding QGP lifetime is about 3.0, 4.5, and 6.5 fm$/c$. To check the accuracy of the estimate based on the relaxation factor, we calculate in the present study the charmonium production rate using the Tsallis distribution for charm and anticharm quarks~\cite{Tsallis:1987eu,Tang:2008ud}. The Tsallis distribution with a parameter $q$ is a generalized Boltzmann distribution, which is the same as the normal Boltzmann distribution when $q=1$ but becomes a harder power law distribution as the value of $q$ increases. In this study, we define $\lambda=q-1$ and determine its values for the initially produced charm and anticharm quarks from hard collisions at RHIC and LHC by fitting the charm quark spectra generated by the event generator PHYTIA~\cite{Sjostrand:2006za}. To study how the charm and anticharm quarks approach equilibrium in a QGP, we use the parton cascade model based on their elastic scattering with light quarks and gluons in the QGP. Fitting these distributions by the Tsallis distribution, we then calculate the charmonium production rate using the transition amplitude that is calculated up to the next-to-leading order (NLO) in perturbative QCD (pQCD)~\cite{Song:2005yd,Park:2007zza}. We find that nonequilibrium charm and anticharm quarks suppress the charmonium production rate compared to that from equilibrated ones and that the suppression factor estimated from the charm quark relaxation time, which has been frequently used in the literature, is close to our results.

This paper is organized as follows. We first give in Sec.~\ref{m2} the transition amplitudes for charmonium production in pQCD. In Sec.~\ref{rates}, we derive the expression for the charmonium production rate from charm and anticharm quarks based on the Tsallis distribution and study the dependence of the $J/\psi$ production rate on the parameters in the Tsallis distribution. We then apply in Sec.~\ref{applications} these results to heavy-ion collisions at RHIC and LHC. Finally, we give the conclusions in Sec.~\ref{conclusions}.  For details on the evaluation of the phase space integral in the charmonium production rate, they are given in Appendix A.

\section{transition amplitudes for charmonium production}\label{m2}

Although the suggestion of $J/\psi$ suppression as a signature of QGP was based on the idea that the $J/\psi$ could not be formed in QGP~\cite{Matsui:1986dk} due to screening of the color charge, recent studies on the spectral functions of charmonia have suggested, on the other hand, that the dissociation temperature of $J/\psi$ is higher than the critical temperature for QGP phase transition~\cite{Hatsuda04,Datta04,Mocsy:2007yj}, indicating that the $J/\psi$ can survive and be regenerated in QGP.

In pQCD, the leading-order process for $J/\psi$ production from charm and anticharm quarks is the reaction $c+\bar{c}\rightarrow J/\psi+g$. The transition amplitude for this reaction is same as that for the $J/\psi$ dissociation reaction $J/\psi+g \rightarrow c+\bar{c}$ given in Refs.~\cite{Song:2005yd,Park:2007zza}, that is
\begin{eqnarray}
|\mathcal{M}|_{\rm LO}^2=4 g^2m_c^2m_{J/\psi}(2k_{0}^2+m_g^2) {\Big|\frac{\partial \psi({\bf
p})}{\partial {\bf p}}\Big|}^2.
\label{LO}
\end{eqnarray}
In the above, $k_0$ and ${\bf p}=\sqrt{m_c(k_0-\epsilon_0)}$ are, respectively, the gluon energy and the relative three momentum of charm and anticharm quarks in the $J/\psi$ rest frame with $\epsilon_0$ being the binding energy of $J/\psi$; $g$ is the strong coupling constant which is taken to be 1.87 from previous phenomenological studies~\cite{Song:2011xi,Song:2011nu}, and $m_c$ and $m_{J/\psi}$ are the masses of charm quark and $J/\psi$, respectively; $m_g$ is the thermal gluon mass obtained from lattice data using the quasiparticle model~\cite{Levai:1997yx}; and $\psi({\bf p})$ is the wavefunction of $J/\psi$. To include the medium effect, we use the temperature-dependent wavefunction and binging energy of $J/\psi$ obtained from solving the Schr\"odinger equation using the screened Cornell potential between charm and anticharm quarks~\cite{Karsch:1987pv}. Eq. (\ref{LO}) becomes the same as that of Bhanot and Peskin~\cite{Peskin:1979va,Bhanot:1979vb} if $\psi({\bf p})$ is taken to be the Coulomb wavefunction for the $1S$ state.

For the next-leading order (NLO) reaction $c+\bar{c}+q(\bar{q},g)\rightarrow J/\psi+q(\bar{q},g)$ in pQCD, its transition amplitude is the same as that for the reaction $J/\psi+q(\bar{q},g) \rightarrow c+\bar{c}+q(\bar{q},g)$. As given in Ref.~\cite{Song:2005yd,Park:2007zza}, the squared transition amplitudes of quark- and gluon-induced NLO processes are
\begin{eqnarray}
|\mathcal{M}|_{\rm qNLO1}^2=8N_c g^4 m_c^2 m_{J/\psi}
{\Big|\frac{\partial \psi({\bf p})}{\partial {\bf p}}\Big|}^2~~~~~~~~~~~~~~~~~~~\nonumber\\
\times\bigg\{-\frac{1}{2}+\frac{k_{10}^2+k_{20}^2}{2 k_1 \cdot k_2}\bigg\},~~~~~~~~~~~~~\label{qNLO1}\\
|\mathcal{M}|_{\rm gNLO1}^2=8(N_c^2-1) g^4 m_c^2 m_{J/\psi}
{\Big|\frac{\partial \psi({\bf p})}{\partial {\bf p}}\Big|}^2~~~~~~~~~~~\nonumber \\
\times\Bigg\{-4+\frac{k_1 \cdot k_2}{k_{10}k_{20}}+\frac{2k_{10}}{k_{20}}+\frac{2k_{20}}{k_{10}}
-\frac{k_{20}^2}{k_{10}^2}-\frac{k_{10}^2}{k_{20}^2}~~~~~\nonumber\\
+\frac{2}{k_1\cdot k_2}\bigg[
\frac{(k_{10}^2+k_{20}^2)^2}{k_{10}k_{20}} -2 k_{10}^2-2
k_{20}^2+k_{10}k_{20}\bigg] \Bigg\},
\label{gNLO1}
\end{eqnarray}
where $N_c$ is the number of quark colors, and $k_1$ and $k_2$ are the momenta of incoming and outgoing partons, respectively.

Other NLO reactions include $c+\bar{c} \rightarrow J/\psi+g+g(q+\bar{q})$ which can be obtained from the reactions $c+\bar{c}+q(\bar{q},g)\rightarrow J/\psi+q(\bar{q},g)$ by changing the incoming parton to an outgoing parton. Their squared amplitudes can thus be obtained from Eqs. (\ref{qNLO1}) and (\ref{gNLO1}) by changing $k_1$ and $k_{10}$ to $-k_1$ and $-k_{10}$, respectively, and by multiplying Eq.~(\ref{qNLO1}) by an overall minus sign and the nubmer of light quark flavors, that is
\begin{eqnarray}
|\mathcal{M}|_{\rm qNLO2}^2=24N_c g^4 m_c^2 m_{J/\psi}
{\Big|\frac{\partial \psi({\bf p})}{\partial {\bf p}}\Big|}^2~~~~~~~~~~~~~~~~~~~\nonumber\\
\times\bigg\{\frac{1}{2}+\frac{k_{10}^2+k_{20}^2}{2 k_1 \cdot k_2}\bigg\},~~~~~~~~~~~~~\label{qNLO2}\\
|\mathcal{M}|_{\rm gNLO2}^2=8(N_c^2-1) g^4 m_c^2 m_{J/\psi}
{\Big|\frac{\partial \psi({\bf p})}{\partial {\bf p}}\Big|}^2~~~~~~~~~~~\nonumber \\
\times\Bigg\{-4+\frac{k_1 \cdot k_2}{k_{10}k_{20}}-\frac{2k_{10}}{k_{20}}-\frac{2k_{20}}{k_{10}}
-\frac{k_{20}^2}{k_{10}^2}-\frac{k_{10}^2}{k_{20}^2}~~~~~\nonumber\\
+\frac{2}{k_1\cdot k_2}\bigg[\frac{(k_{10}^2+k_{20}^2)^2}{k_{10}k_{20}} +2 k_{10}^2+2
k_{20}^2+k_{10}k_{20}\bigg] \Bigg\}.
\label{gNLO2}
\end{eqnarray}

\section{the charmonium production rate}\label{rates}

The production rate of charmonium with momentum ${\bf q}$ in a QGP from the LO and NLO charm and anticharm quark recombination processes described in the previous Section can be written as~\cite{Zhao:2010ti}
\begin{eqnarray}
\frac{dN_{J/\psi}^{\rm LO}}{Vdtd^3{\bf q}}&=&\frac{1}{2E_{\bf q}}\int\prod_{i=1,2}\frac{d^3{\bf p}_i}
{(2\pi)^32E_{p_i}}\int\frac{d^3{\bf k}}{(2\pi)^32E_{k}}\nonumber\\
&\times&(2\pi)^4\delta^{(4)}(p_1+p_2-q-k)\nonumber\\
&\times&f_c({\bf p}_1)f_{\bar{c}}({\bf p}_2)|\mathcal{M}|_{\rm LO}^2,\label{beta1}\\
\frac{dN_{J/\psi}^{\rm NLO1}}{Vdtd^3{\bf q}}&=&\frac{1}{2E_{\bf q}}\int\prod_{i=1,2}\frac{d^3{\bf p}_i}
{(2\pi)^32E_{p_i}}\int\prod_{j=1,2}\frac{d^3{\bf k_j}}{(2\pi)^32E_{k_j}}\nonumber\\
&\times&(2\pi)^4\delta^{(4)}(p_1+p_2-q+k_1-k_2)\nonumber\\
&\times&f_c({\bf p}_1)f_{\bar{c}}({\bf p}_2)f_p(k_{1})|\mathcal{M}|_{\rm NLO1}^2.\label{beta2}\\
\frac{dN_{J/\psi}^{\rm NLO2}}{Vdtd^3{\bf q}}&=&\frac{1}{2E_{\bf q}}\int\prod_{i=1,2}\frac{d^3{\bf p}_i}
{(2\pi)^32E_{p_i}}\int\prod_{j=1,2}\frac{d^3{\bf k_j}}{(2\pi)^32E_{k_j}}\nonumber\\
&\times&(2\pi)^4\delta^{(4)}(p_1+p_2-q-k_1- k_2)\nonumber\\
&\times&f_c({\bf p}_1)f_{\bar{c}}({\bf p}_2)|\mathcal{M}|_{\rm NLO2}^2.\label{beta3}
\end{eqnarray}
In the above, $f_c({\bf p_1})$ and $f_{\bar c}({\bf p_2})$ are, respectively, the charm and anticharm quark distributions; $f_p(k_{1})$ is the thermal parton distribution including the degeneracies of quarks due to their flavors and antiparticles; ${\bf k}$ is the outgoing gluon momentum in the LO $c+\bar{c} \rightarrow J/\psi+g$ reaction; and ${\bf k_{2}}$ is the outgoing parton momentum, while ${\bf k_{1}}$ is the outgoing and incoming parton momentum in the NLO $c+\bar{c}\rightarrow J/\psi+g+g(q+\bar{q})$ and $c+\bar{c}+q(\bar{q},g)\rightarrow J/\psi+q(\bar{q},g)$ reactions, respectively.

\begin{figure}[h]
\centerline{\includegraphics[width=9 cm]{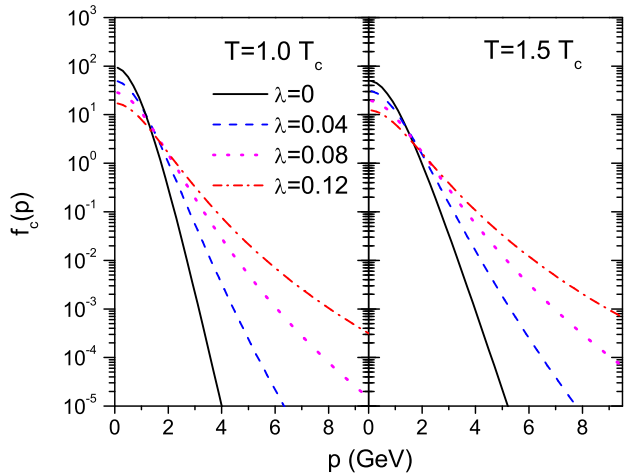}}
\caption{(Color online) Charm quark distribution for different values of $\lambda$ in the Tsallis distribution.}\label{tsallis}
\end{figure}

For the charm and anticharm quark distributions, we use the modified Tsallis distribution function
\begin{eqnarray}
f_{c,{\bar c}}({\bf p})=(2\pi)^3\frac{dN_{c}}{d^3{\bf r}d^3{\bf p}}=A
(\lambda)\left(1+\lambda \frac{E_{\bf p}}{T}\right)^{-1/\lambda},
\label{Tsallis}
\end{eqnarray}
where $\lambda=q-1$ with $q$ being a parameter in the original Tsallis distribution function and $A$, which depends on $\lambda$, is given by
\begin{eqnarray}
A(\lambda)=\frac{2\pi^2n_c}{\int d{\bf p_1} {\bf p_1}^2(1+\lambda E_{\bf p_1}/T)^{-1/\lambda}}
\end{eqnarray}
to ensure that the integration of $f_{c}({\bf p})$ over the charm quark momentum gives the charm quark density $n_c$. In Fig.~\ref{tsallis}, we show the charm quark distribution for different values of $\lambda$. It is seen that as $\lambda$ approaches 0, $f_{c}$ becomes the Boltzmann distribution. Details on the phase-space integrations in Eqs.(\ref{beta1}), (\ref{beta2}), and (\ref{beta3}) with the Tsallis distribution are given in Appendix A.

\begin{figure}[h]
\centerline{\includegraphics[width=9 cm]{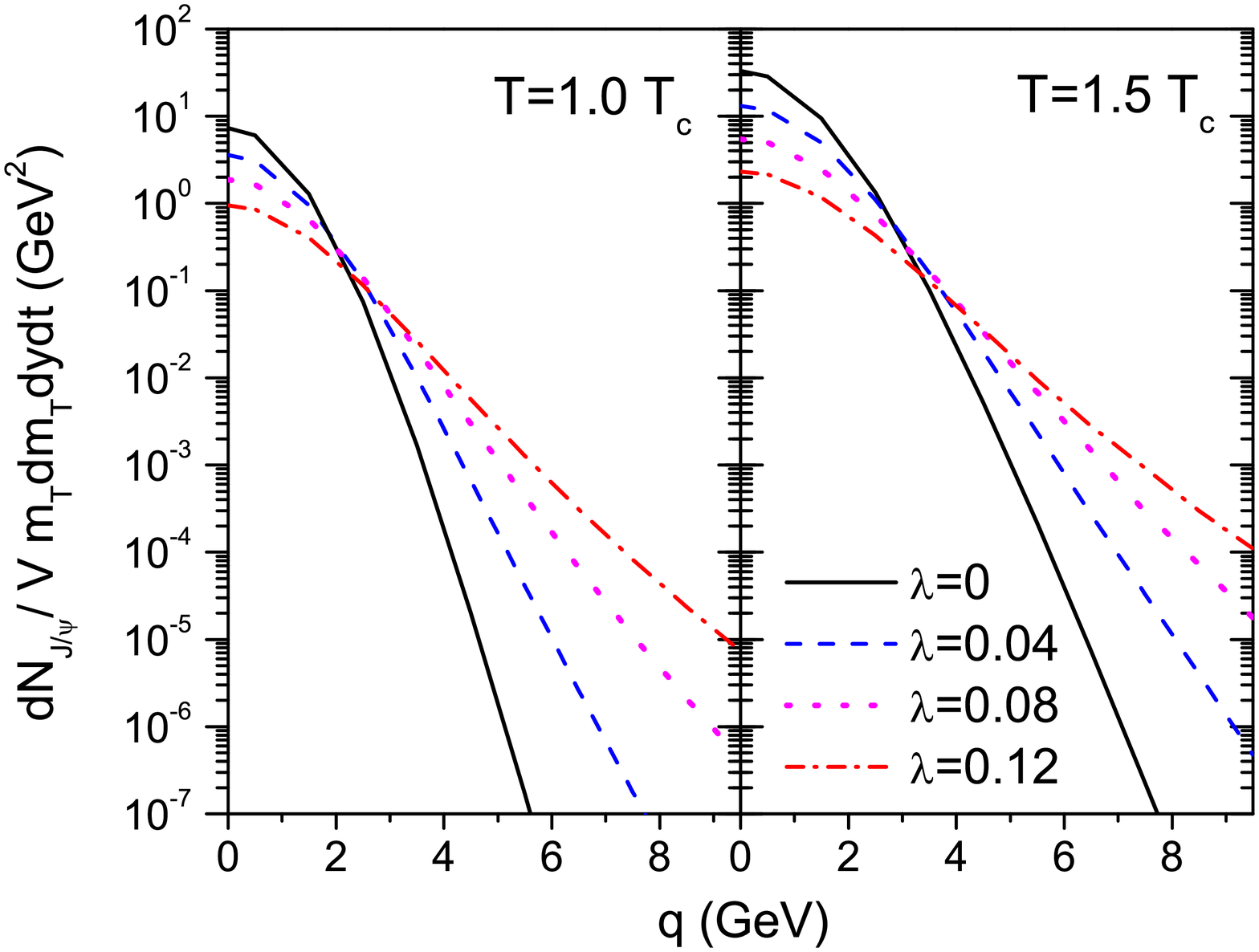}}
\caption{(Color online) Production rate for charmonium of momentum ${\bf q}$ in QGP at temperatures $T=1.0$ (left panel) and $1.5~T_c$ (right panel) from charm and anticharm quarks that have a Tsallis distribution with various values of $\lambda$.}\label{spectra}
\end{figure}

In Fig.~\ref{spectra}, we show the production rate for charmonium of momentum ${\bf q}$ in QGP at temperature $T=1.0$ (left panel) and $1.5~T_c$ (right panel), where $T_c$ is taken to be 170 MeV, from charm and anticharm quarks that have a Tsallis distribution with various values of $\lambda=$0, 0.04, 0.08, and 0.12. These results are obtained with the density of charm quarks in the QGP taken to be 0.008/fm$^3$ (1~$\rm GeV^3$). It is seen that the momentum dependence of the $J/\psi$ production rate becomes harder as $\lambda$ increases. This is reasonable because a larger $\lambda$ indicates a harder charm quark momentum spectrum. Also, the $J/\psi$ production rate is larger at 1.5 $T_c$ than at 1.0 $T_c$, consistent with the larger $J/\psi$ thermal decay width at higher temperature.

\begin{figure}[h]
\centerline{
\includegraphics[width=9 cm]{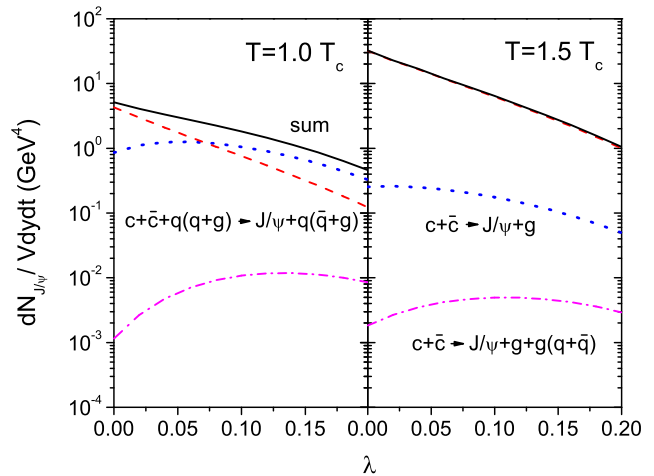}}
\caption{(Color online) Production rate of $J/\psi$ per unit rapidity as a function of $\lambda$ at $T=1.0$ and 1.5 $T_c$. Dotted, dashed, and dot-dashed lines are from the reactions $c+\bar{c}\rightarrow J/\psi+g$, $c+\bar{c}+q(\bar{q},g)\to J/\psi+q(\bar{q},g)$, and $c+\bar{c}\rightarrow J/\psi+g+g(q+\bar{q})$, respectively, and solid lines are their sum.}\label{yields}
\end{figure}

Figure~\ref{yields} shows the dependence of the momentum integrated $J/\psi$ production rate per unit rapidity on the parameter $\lambda$ for the three reactions $c+\bar{c}\to J/\psi+g$ (dotted line), $c+\bar{c}+q(\bar{q},g)\to J/\psi+q(\bar{q},g)$ (dashed line), and $c+\bar{c}\to J/\psi+g+g(q+\bar{q})$ (dot-dashed line). The most dominant process is $c+\bar{c}+q(\bar{q},g)\rightarrow J/\psi+q(\bar{q},g)$, especially at high temperature. The reason for this is that the common factor $|\partial \psi({\bf p})/\partial {\bf p}|^2$ in the squared transition amplitudes in Eq.~(\ref{LO})$-$(\ref{gNLO2}) peaks at small relative momentum ${\bf p}$ between charm and anticharm quarks and the value of ${\bf p}$ is smaller in the reaction $c+\bar{c}+q(\bar{q},g)\rightarrow J/\psi+q(\bar{q},g)$ than in other reactions.

Figure~\ref{yields} also shows that the $J/\psi$ production rate from the three reactions behave differently as the value of $\lambda$ increases. Since harder charm and anticharm quark spectra are less favorable to $J/\psi$ production as a result of the large relative momentum between charm and
anticharm quarks, the production rate of $J/\psi$ from the reaction $c+\bar{c}+q(\bar{q},g)\rightarrow J/\psi+q(\bar{q},g)$ decreases as $\lambda$ increases, similar to that found in the coalescence model \cite{Greco:2003vf}. Although the transition amplitudes in all processes favor soft charm and anticharm quark spectra, to produce additional partons in the final state as in the reactions $c+\bar{c}\rightarrow J/\psi+g$ and $c+\bar{c}\rightarrow J/\psi+g+g(q+\bar{q})$ requires that the initial charm and anticharm quarks to have high energy. As a result, the production rates of $J/\psi$ from the reactions $c+\bar{c}\rightarrow J/\psi+g$ and $c+\bar{c}\rightarrow J/\psi+g+g(q+\bar{q})$ increase and then decrease with increasing $\lambda$, particularly from the latter reaction as it has two thermal partons in the final state. The total $J/\psi$ production rate given by the sum of the production rates from the three reactions decreases, however, with increasing $\lambda$ since it is dominated by the reaction $c+\bar{c}+q(\bar{q},g)\rightarrow J/\psi+q(\bar{q},g)$. Therefore, nonequilibrium charm and anticharm quark distributions suppress the $J/\psi$ production rate in QGP compared to that from completely thermalized charm and anticharm quark distributions.

\section{applications to heavy-ion collisions}\label{applications}

\begin{figure}[h]
\centerline{
\includegraphics[width=9 cm]{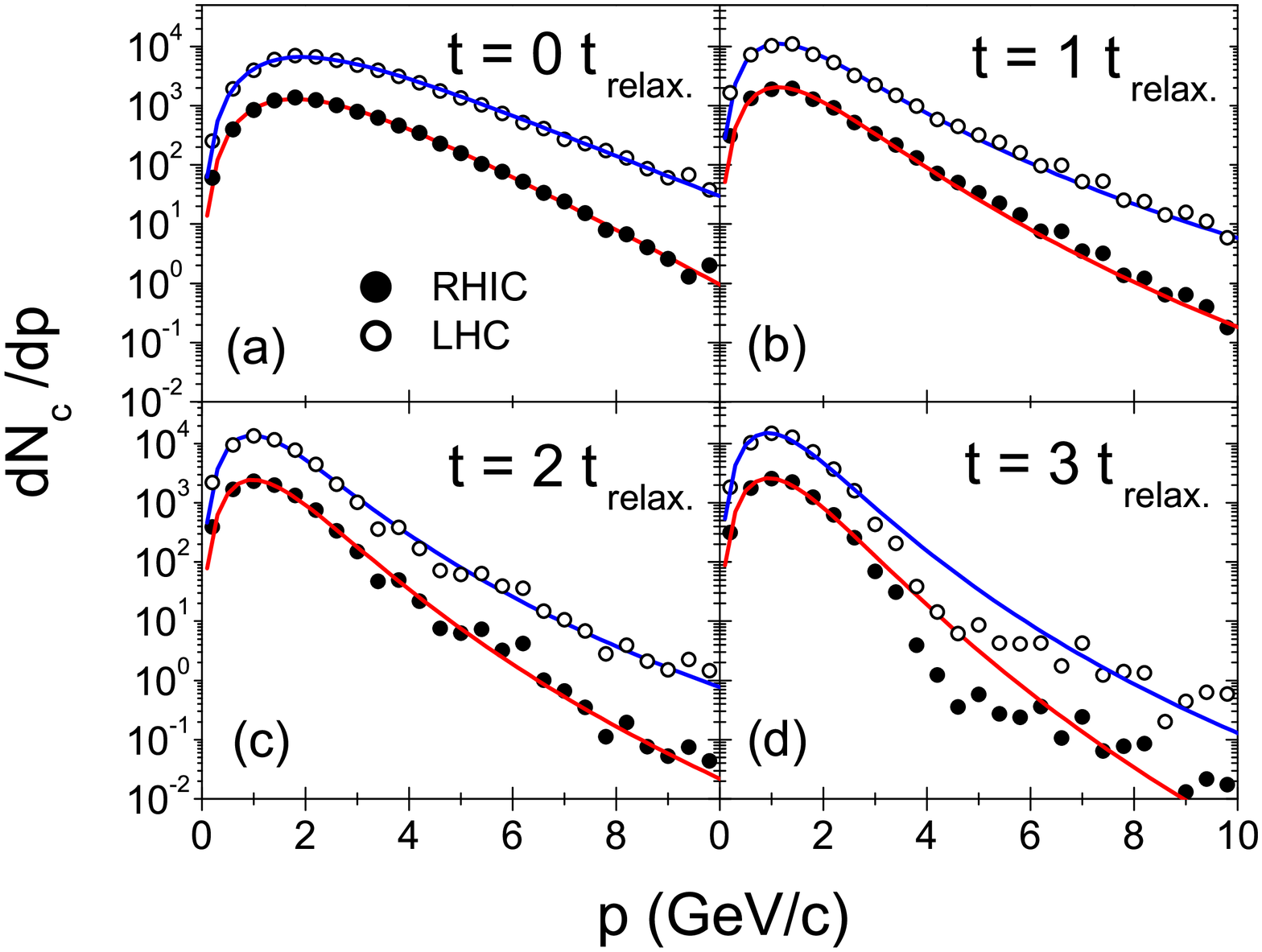}}
\caption{(Color online) Charm quark momentum distribution in arbitrary units for RHIC (solid circles) and for LHC (open circles) in QGP of $T=1.5~T_c$ at 0, 1, 2, and 3 times the charm quark relaxation time $t_{\rm rel}$. Lines are fitted Tsallis distributions with $(T,~\lambda)=$(650~MeV,~0.013), (190~MeV,~0.078), (140~MeV,~0.072), and (150~MeV,~0.058) for RHIC and $(T,~\lambda)=$(680~MeV,~0.033), (140~MeV,~0.107), (90~MeV,~0.097), and (100~MeV,~0.081) for LHC at increasing time.}\label{spectra2}
\end{figure}

To see the relevance of our results to $J/\psi$ production in heavy ion collisions, we carry out a
cascade calculation to study how the charm and anticharm quark distributions change with time in a QGP at a fixed temperature. Specifically, we take the initial charm and anticharm quark spectra at mid-rapidity ($|y|\leq 1$) from the PHYTIA simulations~\cite{Sjostrand:2006za} for p+p collisions at $\sqrt{s}=200~{\rm GeV}$ and $2.76~{\rm TeV}$, available at RHIC and LHC, respectively, and they are shown by solid and open circles in panel (a) of Fig.~\ref{spectra2}. Assuming that the charm and anticharm quark spectra are isotropic in momentum space, we then find that they can be very well described by the Tsallis distribution with $(T,\lambda)=$(650~MeV,~0.013) and (680~MeV,~0.033) as shown, respectively, by lines in panel (a) of Fig.~\ref{spectra2}. Taking the elastic scattering cross sections of charm and anticharm quarks with light quarks and antiquarks to be 1 mb and those with gluons to be 2 mb, we examine the change of their momentum distributions with time. The results for the two initial charm and anticharm quark spectra corresponding to $\sqrt{s}=200~{\rm GeV}$ and $2.76~{\rm TeV}$ in a QGP of temperature $T=1.5~T_c$ are shown by solid and open circles in panels (b)-(d) of Fig.~\ref{spectra2} for different times of 1$t_{\rm rel}$, 2$t_{\rm rel}$, and 3$t_{\rm rel}$. Here $t_{\rm rel}$ denotes the charm (anticharm) quark relaxation time and is defined by
\begin{eqnarray}
t_{\rm rel}^{-1}=\bigg\langle\sum_{i=q,\bar{q},g}\int \frac{d^3k}{(2\pi)^3}n_i(k,T)v_{\rm rel}\sigma_i \bigg(1-\frac{{\bf p}\cdot {\bf p}^\prime}{{\bf p}^2}\bigg)\bigg\rangle,
\end{eqnarray}
where $n_i$ is the density of parton species $i$ including its degeneracy factor in grand canonical ensemble, $v_{\rm rel}$ is the relative velocity between the charm (anticharm) quark and parton, $\sigma_i$ is the elastic cross section of charm (anticharm) quark by the parton, ${\bf p}$ and ${\bf p^\prime}$ are three momenta of incoming and outgoing charm quarks in laboratory frame, respectively, and $\langle\cdots\rangle$ denotes an average over the charm quark momentum distribution. The relaxation time is thus the inverse of drag coefficient of charm quarks~\cite{Svetitsky:1987gq}. We find from the parton cascade simulation that the relaxation time of charm quarks in a QGP of temperature $1.5~T_c$ depends only weakly on their momentum distribution and has an approximate value of about 3.2 fm/$c$. In Fig.~\ref{spectra2}, we also show by lines the fitted charm (anticharm) quark momentum distributions using the Tsallis distribution with the parameters $(T,\lambda)$=(190~MeV,~0.078), (140~MeV,~0.072), and (150~MeV,~0.058) for the case of RHIC and $(T,\lambda)$=(140~MeV,~0.107), (90~MeV,~0.097), and (100~MeV,~0.081) for the case of LHC at different times of 1$t_{\rm rel}$, 2$t_{\rm rel}$, and 3$t_{\rm rel}$. It is seen that the Tsallis distribution again describes very well the charm (anticharm) quark distribution from the cascade simulation. We note that the results shown in Fig.~\ref{spectra2} with the time measured in unit of the relaxation time hardly depends on the value of elastic cross sections of charm and anticharm quarks.

\begin{figure}[h]
\centerline{
\includegraphics[width=9 cm]{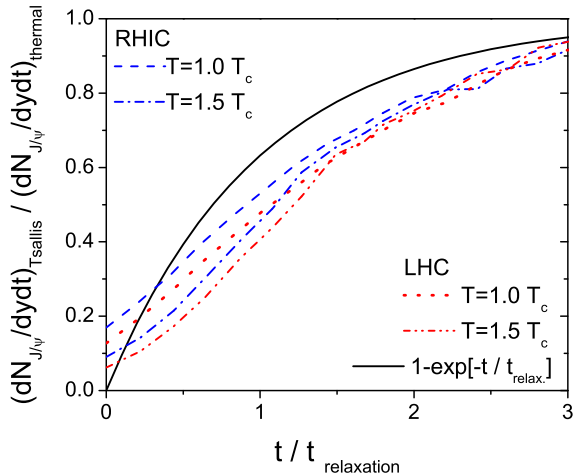}}
\caption{(Color online) Time dependence of the ratio of the $J/\psi$ production rate from nonequilibrium charm and anticharm quarks to that from completely thermalized charm and anticharm quarks in QGP at $T=$1.0 and 1.5 $T_c$. The solid line is the result based on the relaxation factor correction.}
\label{relaxation}
\end{figure}

Using the $J/\psi$ production rate given in Fig.~\ref{yields} and the charm and anticharm quark distribution given in Fig.~\ref{spectra2}, we can calculate the $J/\psi$ production rate and obtain its ratio with respect to the rate from completely thermalized charm and anticharm quarks. This is shown in Fig.~\ref{relaxation} as a function of time for a QGP of temperature $T=$1.0 and 1.5 $T_c$ in collisions at RHIC and LHC. These results show that the suppression of $J/\psi$ production due to the nonequilibrium charm and anticharm quark distributions is stronger at higher-energy collisions and at higher temperature. This is due to the fact that the more energetic charm and anticharm quarks produced in higher energy collisions make it harder to produce the charmonium, and that the smaller charmonium binding energy at higher temperature further suppresses its production. Although the suppression starts at different values in different cases, their time dependences all seem to be similar. Since the temperature of QGP formed in relativistic heavy ion collisions decreases with time, the suppression of $J/\psi$ production due to nonequilibrium charm anticharm quarks would decrease with time as well. For comparison, we also show in Fig.~\ref{relaxation} by the solid line the results from using the relaxation factor $R\equiv 1-\exp[-\tau/\tau_{\rm rel}]$ that has been used in the literature to take into account the effect of nonequilibrium charm and anticharm quark distributions~\cite{Zhao:2007hh,Song:2010ix,Song:2011xi}. It is seen that the suppression of the $J/\psi$ production rate based on the nonequilibrium Tsallis distribution is similar to the relaxation factor correction. Assuming the lifetime of QGP is given by the relaxation time of charm quarks, the $J/\psi$ production rates in heavy ion collisions at RHIC and LHC are then suppressed, respectively, by 40-50 \% and 35-45 \% compared to those from completely thermalized charm and anticharm quarks.

\section{conclusions}\label{conclusions}

Using the transition amplitudes for $J/\psi$ production from charm and anticharm quarks calculated up to NLO in pQCD and including the medium effect on the $J/\psi$ wavefunctions and binding energies from a screened Cornell potential between the charm and anticharm quarks in QGP, we have studied charmonium production in QGP from nonequilibrium charm and anticharm quarks that are described by the Tallis distribution. We have found that nonequilibrium charm and anticharm quarks suppress the production rate of $J/\psi$ in QGP, compared to the rate from completely thermalized charm and anticharm quarks. We have further used the calculated $J/\psi$ production rate in QGP to study $J/\psi$ production in heavy-ion collisions at RHIC and LHC by using the PHYTIA simulation to obtain the initial charm and anticharm quark distributions and then follow their elastic scattering with light quarks and antiquarks as well as gluons in QGP via the cascade simulation. With the resulting charm and anticharm quark distribution parameterized by the Tsallis distribution, we have found that the suppression in the $J/\psi$ production rate is stronger in higher energy collisions and at higher temperatures, although their time dependence is similar in all cases. We have also compared our results to those in the literature obtained with the relaxation factor correction using the charm quark relaxation time and found that the latter is similar to our results.

\section*{Acknowledgements}

We thank Rapp Ralf for helpful comments on the charm quark relaxation time. This work was supported in part by the U.S. National Science Foundation under Grant Nos. PHY-0758115 and PHY-1068572, the US Department of Energy under Contract No. DE-FG02-10ER41682, and the Welch Foundation under Grant No. A-1358.

\appendix
\section{}

In this Appendix, we give details on the evaluation of the $J/\psi$ production rates in QGP from the reactions $c+\bar{c}\rightarrow J/\psi+g$, $c+\bar{c}\rightarrow J/\psi+g+g(q+\bar{q})$, and $c+\bar{c}+q(\bar{q},g)\rightarrow J/\psi+q(\bar{q},g)$, based on the Tsallis distribution for the charm and anticharm quark distributions.

\subsection{$c+\bar{c}\rightarrow J/\psi+g$}

Substituting Eq.~(\ref{Tsallis}) into Eq.~(\ref{beta1}) and using $d^3{\bf q}=E_{\bf q}m_Tdm_Tdy$ in terms of the $J/\psi$ transverse mass $m_T$ and rapidity $y$, we obtain
\begin{eqnarray}
\frac{dN_{J/\psi}^{\rm LO}}{m_Tdm_Tdydt}=\frac{A^2(\lambda)V}{8\pi^2}\int\frac{d^3{\bf p_1}}{(2\pi)^32E_{\bf p_1}}\int\frac{d^3{\bf p}_2}{(2\pi)^32E_{\bf p_2}}\quad\nonumber\\
\times\int\frac{d^3{\bf k}}{(2\pi)^32E_{\bf k}}\bigg(1+\lambda \frac{E_{\bf p_1}}{T}\bigg)^{-1/\lambda}\bigg(1+\lambda \frac{E_{\bf p_2}}{T}\bigg)^{-1/\lambda}\nonumber\\
\times(2\pi)^4\delta^4(p_1+p_2-q-k)|\mathcal{M}|_{\rm LO}^2,\qquad\qquad\qquad\quad~
\end{eqnarray}
where $V$ is the volume of the system. To evaluate above integrals, we first express the Tsallis  distribution function in covariant form by replacing $E_{\bf p}$ with $p\cdot u$, where $u^\mu=\gamma(1,\vec{\beta})$ with $\beta$ being the boost velocity and $\gamma=(1-\beta^2)^{-1/2}$. Transforming to the center of mass frame of charm and anticharm quarks via ${\bf P}={\bf p_1}+{\bf p_2}$ and ${\bf p}=({\bf p_1}-{\bf p_2})/2$, which leads to $\beta={\bf P}/E$ and $\gamma=E/\sqrt{s}$, with $E=E_{\bf p_1}+E_{\bf p_2}$ and $s=(q+k)^2=(p_1+p_2)^2=E^2-{\bf P}^2$, we then have
\begin{eqnarray}
\frac{dN_{J/\psi}^{\rm LO}}{m_Tdm_Tdydt}=\frac{A^2(\lambda)V}{8\pi^2}\int\frac{d^3{\bf k}}{(2\pi)^32E_{\bf k}}\int\frac{d^3{\bf p}}{(2\pi)^2 s}~~~~\nonumber\\
\times\bigg[\Big(1+\frac{\lambda\gamma\sqrt{s}}{2T}\Big)^2-\Big(\frac{\lambda\gamma \vec{\beta}\cdot{\bf p}}{T}\Big)^2\bigg]^{-1/\lambda}\nonumber\\
\times \delta(\sqrt{s}-E_{\bf q}-E_{\bf k})|\mathcal{M}|_{\rm LO}^2~~~~~~~~~~~~~~~\nonumber\\
\nonumber\\
=\frac{A^2(\lambda)V}{16(2\pi)^3}\int\frac{d^3{\bf k}}{(2\pi)^32E_{\bf k}}\sqrt{1-\frac{4m_c^2}{s}}~|\mathcal{M}|_{\rm LO}^2\int_{-1}^1d\cos\theta~\nonumber\\
\times \bigg[\Big(1+\frac{\lambda\gamma\sqrt{s}}{2T}\Big)^2-\Big(1-\frac{4m_c^2}{s}\Big)\Big(\frac{\lambda {\bf P}}{2T}\Big)^2\cos^2\theta\bigg]^{-1/\lambda},~
\end{eqnarray}
where $\theta$ is the angle between ${\bf p}$ and $\vec{\beta}$.

Choosing the momentum ${\bf q}$ of produced $J/\psi$ to be the $z$-axis, i.e., $q^\mu=(E_{\bf q}, 0, 0, q)$, and denoting $k^\nu=(E_{\bf k}, 0, k\sin\phi, k\cos\phi)$, we then have
\begin{eqnarray}
\frac{dN_{J/\psi}^{\rm LO}}{m_Tdm_Tdydt}=\frac{A^2(\lambda)V}{32(2\pi)^5}\int_0^\infty\frac{d{k}{k}^2}{E_{\bf k}}\int_{-1}^1d\cos\phi~~~~~~~~~~~~~~\nonumber\\
\times\sqrt{1-\frac{4m_c^2}{s}}~|\mathcal{M}|_{\rm LO}^2\int_{-1}^1d\cos\theta~~~~~~~~~~~~~~~~~\nonumber\\
\times \left[\left(1+\frac{\lambda\gamma\sqrt{s}}{2T}\right)^2-\left(1-\frac{4m_c^2}{s}\right)\left(\frac{\lambda {\bf P}}{2T}\right)^2\cos^2\theta\right]^{-1/\lambda},\nonumber\\
\end{eqnarray}
where ${\bf P}^2={\bf q}^2+{\bf k}^2+2qk\cos\phi$.
We note that the squared transition amplitude, Eq.~(\ref{LO}), is a function of $k_0$, which is the energy of thermal gluon in the $J/\psi$ rest frame and can be expressed as $k_0=(s-m_{J/\psi}^2-m_g^2)/2m_{J/\psi}$.

\subsection{$c+\bar{c}+q(\bar{q},g)\rightarrow J/\psi+q(\bar{q},g)$}

As in the previous case, the production rate from this reaction can be expressed in the center of mass frame of charm and anticharm quarks, and it is given by
\begin{eqnarray}
\frac{dN_{J/\psi}^{\rm NLO1}}{m_Tdm_Tdydt}=\frac{A^2(\lambda)V}{16(2\pi)^3}\int\frac{d^3{\bf k}_1}{(2\pi)^32E_{\bf k_1}}\int\frac{d^3{\bf k}_2}{(2\pi)^32E_{\bf k_2}}~~~\nonumber\\
\times\sqrt{1-\frac{4m_c^2}{w}}~|\mathcal{M}|_{\rm NLO1}^2 f({\bf k}_1)\int_{-1}^1d\cos\theta~~~~~~~~~~~~~~~\nonumber\\
\times \bigg[\Big(1+\frac{\lambda\gamma\sqrt{w}}{2T}\Big)^2-\Big(1-\frac{4m_c^2}{w}\Big)\Big(\frac{\lambda {\bf P}}{2T}\Big)^2\cos^2\theta\bigg]^{-1/\lambda},~
\end{eqnarray}
where ${\bf k}_1$ and ${\bf k}_2$ are, respectively, the three momenta of incoming and outgoing partons; ${\bf P}={\bf p}_1+{\bf p}_2$ and $w=(p_1+p_2)^2$; and $f({\bf k}_1)$ is the Fermi-Dirac distribution in the quark-induced reaction and the Bose-Einstein distribution in the gluon-induced reaction. Furthermore, $f({\bf k}_1)$ should be multiplied by the flavor number in the case of quark-induced reaction.

Choosing the $z$-axis along the $J/\psi$ momentum and using
\begin{eqnarray}
k_1^\nu&=&(E_{\bf k_1}, 0, {k}_1\sin\phi, {k}_1\cos\phi),\nonumber\\
k_2^\lambda&=&(E_{\bf k_2}, {k}_2\sin\theta_1\cos\theta_2, {k}_2\sin\theta_1\sin\theta_2, {k}_2\cos\theta_1),\nonumber\\
P^\sigma&=&(E_{\bf q}+E_{\bf k_1}-E_{\bf k_2},~-{k}_2\sin\theta_1\cos\theta_2,\nonumber\\
&&\hspace{-0.5cm}{k}_1\sin\phi-{k}_2\sin\theta_1\sin\theta_2,~{k}_1\cos\phi-{k}_2\cos\theta_1),
\end{eqnarray}
where $P=p_1+p_2$, the production rate becomes
\begin{eqnarray}
\frac{dN_{J/\psi}^{\rm NLO1}}{m_Tdm_Tdydt}=\frac{A^2(\lambda)V}{64(2\pi)^{8}}\int_a^\infty\frac{d{k}_1{k}_1^2}{E_{\bf k_1}}\int_{-1}^b d\cos\phi~~~~~~\nonumber\\
\times \int\frac{d{k}_2{k}_2^2}{E_{\bf k_2}}\int d\cos\theta_1\int d\theta_2~~~~~~~~~~~~~~\nonumber\\
\times\sqrt{1-\frac{4m_c^2}{w}}~|\mathcal{M}|_{\rm NLO1}^2\int_{-1}^1d\cos\theta~~~~~~~~~~~~~~\nonumber\\
\times \bigg[\Big(1+\frac{\lambda\gamma\sqrt{w}}{2T}\Big)^2-\Big(1-\frac{4m_c^2}{w}\Big)\Big(\frac{\lambda {\bf P}}{2T}\Big)^2\cos^2\theta\bigg]^{-1/\lambda},
\end{eqnarray}
where
\begin{eqnarray}
{\bf P}^2&=&{\bf q}^2+{\bf k}_1^2+{\bf k}_2^2+2{q}({k}_1\cos\phi-{k}_2\cos\theta_1)\nonumber\\
&&-2{k}_1{k}_2(\cos\phi\cos\theta_1+\sin\phi\sin\theta_1\sin\theta_2),\nonumber\\
w&=&(E_{\bf q}+E_{\bf k_1}-E_{\bf k_2})^2-{\bf P}^2,\nonumber\\
s&=&m_{J/\psi}^2+m_{q(\bar{q},g)}^2+2E_{\bf q}E_{\bf k_1}-2{q}{k}_1\cos\phi.
\end{eqnarray}

Because the incoming energy is always larger than the invariant mass of final state, i.e., $s\geq(2m_c+m_{q(\bar{q},g)})^2$, ${k}_1$ and $\cos\phi$ have the following integration limits
\begin{eqnarray}
a&=&\frac{-C+\sqrt{C^2-4BD}}{2B},\nonumber\\
b&=&\frac{m_{J/\psi}^2-4m_c^2+4m_cm_{q(\bar{q},g)}+2E_{\bf q}E_{\bf k_1}}{2{q k_1}},
\end{eqnarray}
with
\begin{eqnarray}
B&=&4m_{J/\psi}^2,\nonumber\\
C&=&4{q}(4m_c^2-m_{J/\psi}^2+4m_cm_{q(\bar{q},g)}),\nonumber\\
D&=&4E_{\bf q}^2m_{q(\bar{q},g)}^2-(4m_c^2-m_{J/\psi}^2+4m_cm_{q(\bar{q},g)})^2.\nonumber
\end{eqnarray}
For the integration ranges of ${k}_2$, $\cos\theta_1$, and $\theta_2$, they are such that $w\geq 4m_c^2$.

The squared transition amplitude for this reaction, i.e., Eq.~(\ref{qNLO1}) or (\ref{gNLO1}), is a function of the energies $k_{10}$ and $k_{20}$ of incoming and outgoing partons given by
\begin{eqnarray}
k_{10}=\frac{s-m_{J/\psi}^2-m_g^2}{2m_{J/\psi}},\quad
k_{20}=\frac{m_{J/\psi}^2+m_g^2-v}{2m_{J/\psi}},
\end{eqnarray}
with
\begin{eqnarray}
v&=&(q-k_2)^2\nonumber\\
&=&m_{J/\psi}^2+m_{q(\bar{q},g)}^2-2E_{\bf q}E_{\bf k_2}+2{q}{k}_2\cos\theta_1,\nonumber\\
\end{eqnarray}
and of
\begin{eqnarray}
k_1\cdot k_2=E_{\bf k_1}E_{\bf k_2}~~~~~~~~~~~~~~~~~~~~~~~~~~~~~~~~~~\nonumber\\
-{k}_1{k}_2(\cos\phi\cos\theta_1+\sin\phi\sin\theta_1\sin\theta_2).
\end{eqnarray}

\subsection{$c+\bar{c}\rightarrow J/\psi+g+g(q+\bar{q})$}

The production rate for this reaction is similar to the one for the reaction $c+\bar{c}\rightarrow J/\psi+g$, i.e.,
\begin{eqnarray}
\frac{dN_{J/\psi}^{\rm NLO2}}{m_Tdm_Tdydt}=\frac{A^2(\lambda)V}{16(2\pi)^3}\int\frac{d^3{\bf k}_1}{(2\pi)^32E_{\bf k_1}}\int\frac{d^3{\bf k}_2}{(2\pi)^32E_{\bf k_2}}~~~\nonumber\\
\times\sqrt{1-\frac{4m_c^2}{s}}~|\mathcal{M}|_{\rm NLO2}^2\int_{-1}^1d\cos\theta~~~~~~~~~~~~~~~\nonumber\\
\times \bigg[\Big(1+\frac{\lambda\gamma\sqrt{s}}{2T}\Big)^2-\Big(1-\frac{4m_c^2}{s}\Big)\Big(\frac{\lambda {\bf P}}{2T}\Big)^2\cos^2\theta\bigg]^{-1/\lambda},~
\end{eqnarray}
where $k_1$ and $k_2$ are momenta of two outgoing thermal partons.

Again, choosing the $z$-axis along the $J/\psi$ momentum and using
\begin{eqnarray}
k_1^\nu=(E_{\bf k_1}, 0, {\bf k}_1\sin\phi, {\bf k}_1\cos\phi),~~~~~~~~~~~~~~~~~~~~~~\nonumber\\
k_2^\lambda=(E_{\bf k_2}, {\bf k}_2\sin\theta_1\cos\theta_2, {\bf k}_2\sin\theta_1\sin\theta_2, {\bf k}_2\cos\theta_1),
\nonumber\\
\end{eqnarray}
the $J/\psi$ production rate is then
\begin{eqnarray}
\frac{dN_{J/\psi}^{\rm NLO2}}{m_Tdm_Tdydt}=\frac{A^2(\lambda)V}{64(2\pi)^{8}}\int_0^\infty\frac{d{\bf k}_1{\bf k}_1^2}{E_{\bf k_1}}\int_{-1}^1d\cos\phi~~~~~~\nonumber\\
\times \int_0^\infty\frac{d{\bf k}_2{\bf k}_2^2}{E_{\bf k_2}}\int_{-1}^1d\cos\theta_1\int_0^{2\pi}d\theta_2~~~~~~~~~~~~~~\nonumber\\
\times\sqrt{1-\frac{4m_c^2}{s}}~|\mathcal{M}|_{\rm NLO2}^2\int_{-1}^1d\cos\theta~~~~~~~~~~~~~~\nonumber\\
\times \bigg[\Big(1+\frac{\lambda\gamma\sqrt{s}}{2T}\Big)^2-\Big(1-\frac{4m_c^2}{s}\Big)\Big(\frac{\lambda {\bf P}}{2T}\Big)^2\cos^2\theta\bigg]^{-1/\lambda},
\end{eqnarray}
where
\begin{eqnarray}
{\bf P}^2&=&{\bf q}^2+{\bf k}_1^2+{\bf k}_2^2+2{q}({k}_1\cos\phi+{k}_2\cos\theta_1)\nonumber\\
&&+2{k}_1{k}_2(\cos\phi\cos\theta_1+\sin\phi\sin\theta_1\sin\theta_2),\nonumber\\
s&=&(E_{\bf q}+E_{\bf k_1}+E_{\bf k_2})^2-{\bf P}^2.\nonumber
\end{eqnarray}

The squared transition amplitude, i.e., Eq.~(\ref{qNLO2}) or (\ref{gNLO2}), is a function of the energies $k_{10}$ and $k_{20}$ of thermal partons in the $J/\psi$ rest frame,
\begin{eqnarray}
k_{10}=\frac{s_1-m_{J/\psi}^2-m_g^2}{2m_{J/\psi}},\quad
k_{20}=\frac{s_2-m_{J/\psi}^2-m_g^2}{2m_{J/\psi}},
\end{eqnarray}
where
\begin{eqnarray}
s_1&=&(q+k_1)^2\nonumber\\
&=&(E_{\bf q}+E_{\bf k_1})^2-{\bf q}^2-{\bf k}_1^2-2{qk}_1\cos\phi,\nonumber\\
s_2&=&(q+k_2)^2\nonumber\\
&=&(E_{\bf q}+E_{\bf k_2})^2-{\bf q}^2-{\bf k}_2^2-2{qk}_2\cos\theta_1,\nonumber
\end{eqnarray}
and of
\begin{eqnarray}
k_1\cdot k_2=\frac{s-s_1-s_2+m_{J/\psi}^2}{2}.
\end{eqnarray}


\end{document}